\documentclass[
10pt,tightenlines,
amsmath,
amssymb,
nofootinbib,
prc,
twocolumn,
showpacs,
preprintnumbers]{revtex4}

\usepackage{graphicx}
\usepackage{psfrag}

\newcommand{\beq}{\begin{equation}}
\newcommand{\eeq}{\end{equation}}
\newcommand{\bea}{\begin{eqnarray}}
\newcommand{\eea}{\end{eqnarray}}
\newcommand{\bmat}{\begin{pmatrix}}
\newcommand{\emat}{\end{pmatrix}}
\newcommand{\nn}{\nonumber}
\newcommand{\junk}[1]{}
\def\<{\langle}
\def\>{\rangle}
\def\d{\partial}
\def\+{\dagger}

\def\UEM{$U(1)_{EM}~$}
\def\rtil{\tilde{r}}

\def\da{\delta a}

\def\pvev{\< |\psi_1|^2 \>}
\def\nvev{\< |\psi_2|^2 \>}
\def\mcoop{m_c}

\begin{document}

\title{Vortices and type-I superconductivity in neutron stars}
\author{Kirk~B.~W.~Buckley, Max~A.~Metlitski, and Ariel~R.~Zhitnitsky}
\affiliation{Department of Physics and Astronomy, \\
University of British Columbia, \\
Vancouver, BC, Canada, V6T 1Z1  }
\begin{abstract}
In a recent paper by Link, it was pointed out that the
the standard picture of the neutron star core, composed of a mixture
of a neutron superfluid and a proton type-II superconductor, is
inconsistent with observations of long period precession in isolated
pulsars. In the following we will show that intervortex force between
the magnetic flux tubes may be attractive resulting in a type-I
(rather than type-II) superconductor. In this case the
magnetic field cannot exist in the form of magnetic flux tubes,
supporting Link's observation. This behavior of the system is due to the
strong interaction between the proton-neutron Cooper pairs, which was
previously ignored.
We also calculate the critical magnetic fields $H_c$ and $H_{c2}$
for type-I/II superconductors. These results also support our claim
of type-I superconductivity in the cores of neutron stars.

\end{abstract}

\maketitle

\section{Introduction}

Recently, it was pointed out by Link \cite{link} that observations of
long period precession \cite{precession} may be inconsistent with the
standard picture of the interior of a neutron star. In the
conventional picture, the extremely dense interior is
mainly composed of neutrons, with a small amount of protons and electrons
in beta equilibrium. The
neutrons form $^3P_2$ Cooper pairs and Bose condense to a superfluid
state, while the protons form $^1S_0 $ Cooper pairs and Bose condense
to give a superconductor as well
(see e.g. Ref. \cite{review} for a recent review).
It is generally assumed that
the proton superconductor is a type-II superconductor, which means
that it supports a stable lattice of magnetic flux tubes
in  the presence of a magnetic field. This belief
is based on  simple estimations of the coherence length and the London
penetration depth which  ambiguously imply a type-II superconductivity.
In addition, the
rotation of a neutron star causes a lattice of quantized vortices
to form in the superfluid neutron state, similar to the observed
vortices  that form when  superfluid $He$ is rotated fast enough.
The axis of rotation and the axis the magnetic field are not aligned.
This, coupled with the fact that the two different types of vortices
interact quite strongly lead to the suggestion that the
observed long period precession may require reexamining the picture of
type-II superconductivity inside neutron stars \cite{link} that follows
from the standard analysis when only a single proton field is considered.
A possible resolution of this ``apparent contradiction" was
suggested in our recent letter \cite{typeI}, where we argued
that if one takes into account that the Cooper pairs of neutrons
are also present in the system and they strongly interact with
proton Cooper pairs, the superconductor may in fact be type-I,
even when a naive analysis that only includes the proton degrees of
freedom  seems to indicate type-II behavior.
If this scenario is realized in nature, this means that the interior
of neutron stars would exhibit the Meissner effect
and therefore would not support a
stable lattice of magnetic flux tubes. This would resolve the
apparent discrepancy \cite{link} between the observation of long period
precession \cite{precession} and the typical parameters of the neutron
stars which naively suggest
type-II superconductivity in neutron stars. The main goal of this paper is to
provide a  detailed analysis   with complete calculations
supporting the claim  of  our short letter \cite{typeI}.

It is well known that type-II superconductors have magnetic flux tubes
in the presence of a magnetic field. In the interior of neutron stars,
which is a mixture of neutron and proton superfluids, the proton
superfluid will form a vortex lattice of magnetic flux tubes if the
superconductor is type-II. Inside the core of these vortices, the proton
condensate vanishes, and the core is filled with normal protons
resulting in the restoration of the broken \UEM symmetry. For
accepted estimates of proton correlation length and London
penetration depth, the distant proton vortices repel each other
leading to formation of a stable vortex lattice. This is the
standard picture realized in conventional type-II superconductors.
However, there are many situations where this picture will
be qualitatively modified. For example, if there is a second component
(such as a neutron
component in our specific case), it may be energetically favorable for
the cores of vortices to be filled with a nonzero condensate of this
second component, as it was originally suggested
in the cosmological context by Witten \cite{witten,defectsbook}.
There are numerous
examples of physical systems where this phenomena occurs: superconducting
cosmic strings in cosmology, magnetic flux tubes in the high $T_c$
superconductors, Bose-Einstein condensates, superfluid $~^3He$, and
high baryon density quark matter
\cite{witten,defectsbook,volovik,BEC,highTc,krstrings,superk}.
Given this, one might guess that such nontrivial vortex structure
may occur in the core of neutron stars where we have another
example of a two component system.
We shall argue in what follows that if the interaction
between proton and neutron Cooper pairs is approximately
equal (a precise  condition of this ``approximately" will be derived below),
the vortex-vortex interaction will be modified and the system
will be a type-I superconductor where the magnetic field is completely
expelled from the bulk.\footnote{In reality, the magnetic field
must be present in the neutron
star interior. The way this picture may be realized in nature is through
the formation of domains of superconductor matter and normal matter.}
The main assumption  that we are making is that
the interactions between the proton and neutron Cooper pairs are
approximately equal, leading to an approximate $U(2)$ symmetry. We
believe that this assumption is justified by the original isospin $SU(2)$
symmetry of the   neutrons and protons. The result of this is that
the proton vortices or magnetic flux tubes have nontrivial core structure.
The superfluid density of the neutrons is larger in the vortex core than
at spatial infinity. In addition, the size of the vortex core and the
asymptotic behavior of the proton condensate is modified due to the
additional neutron condensate. The most important result of these
effects is that the interaction between distant proton vortices
may be attractive in a physically realizable region of parameter
space leading to type-I behavior: destruction of the
proton vortex lattice and expulsion of the magnetic flux from the
superconducting region of the neutron star.

This paper is organized as follows. In Sec. II we will introduce the
free energy that describes a two component (neutron-proton)
superfluid/superconducting system and show that the addition of the
second component (neutron) leads to nontrivial core structure, which
alters the properties of the magnetic flux tube. In Sec. III, we will
give two different calculations of the interaction between two widely
separated vortices. These two different methods lead to the same
conclusion, that the interior of a neutron star may be a type-I
superconductor. In the subsequent section, Sec. IV, we will calculate
the critical magnetic fields $H_c$ and $H_{c2}$ (the magnetic fields
above which superconductivity is destroyed in type-I and type-II
superconductors,
respectively). These results confirm our findings in Sec. III that type-I
superconductivity may occur in the interior of neutron stars.
In Sec. V we will end with concluding remarks on possible implications
of our results. Specifically, we will comment on the possible nature of
glitches (observed in many systems) when  the environment of the
neutron star core is a type-I superconductor.

\section{Structure of Magnetic Flux Tubes}

We start by considering the following effective Landau-Ginsburg
free energy that describes a two component superfluid Bose
condensed system. In our system, we have a proton condensate
described by $\psi_1$ and a neutron condensate described by
$\psi_2$. We do not consider the normal component of the protons
and neutrons with their specific excitations, only the superfluid
component. The $\psi_1$ field with electric charge $q$ (which is
actually twice the fundamental charge of the proton, $q=2|e|$)
interacts with the gauge field ${\bf A}$, with ${\bf B} =\nabla
\times {\bf A}$. The two dimensional free energy reads (we neglect
the dependence on third direction along the vortex such that
${\cal F}$ measures the free energy per unit length): \bea
\label{dimfree} {\cal F} &=& \int d^2 r
    \left( \right.
    \frac{\hbar^2}{2 \mcoop}(|(\nabla - \frac{i q}{\hbar c}{\bf A})\psi_1|^2
    + |\nabla \psi_2|^2)    \nn \\
    &+& \frac{{\bf B}^2}{8 \pi} + V(|\psi_1|^2,|\psi_2|^2) \left. \right)
\eea
where $\mcoop = 2 m$ and $m$ is the mass of the nucleon.
Here we have moved the effective mass difference of the proton and
neutron Cooper pairs onto the interaction potential $V$. In the
free energy given above, we have ignored the term coupling the proton
and neutron superfluid velocities, which gives rise to the
Andreev-Bashkin effect \cite{Bashkin}, as it is not important in
our discussion. We have also ignored the fact that the neutron
condensate has a non-trivial $^3P_2$ order parameter as only the
magnitude of the neutron condensate is relevant to the effect
described below.

The free energy (\ref{dimfree}) is invariant under a $U(1)_1 \times
U(1)_2$ symmetry associated with respective phase rotations of
fields $\psi_1$ and $\psi_2$, which corresponds to the
conservation of the number of Cooper pairs for each species of
particles.
Moreover, we assume, that the proton and neutron Cooper pairs
interact approximately in the same way. Therefore, the interaction
potential $V$ can be approximately written as
$V(|\psi_1|^2,|\psi_2|^2) \approx U(|\psi_1|^2 + |\psi_2|^2)$.

Naively, one might think that such an assumption can not
be justified due to the well-known differences in structure
of the Fermi-surfaces and gaps (which strongly depend
on the Fermi energies) of neutron and proton Cooper pairs in neutron stars.
 However, this does not imply that the   interaction at large distances
(much larger  than the inverse gap) between proton  and neutron Cooper pairs
(not between protons and neutrons) is very different or  that their
respective Bose chemical potentials $\mu_i$ (do not confuse with original
Fermi chemical potentials corresponding to protons and neutrons) are very
different. We expect that at very large distances which are relevant for
the analysis of  the vortex-vortex  interaction,
the internal structure of the gap as well as the differences in the densities
of proton and neutron Cooper pairs $n_{1,2}$ (do not confuse with
proton and neutron densities) are not very important provided that
scattering lengths   of different  Cooper pair species  are approximately
the same. In the case where $m_d=m_u$ and  electromagnetic interactions are
neglected, we expect that the scattering lengths for proton and neutron Cooper
pairs would be exactly the same. When $m_d\neq m_u$ we expect that
the effect of the asymmetry due to the renormalization with the environment
would be  proportional to $m_d-m_u$. In principle,
some mild singularity such as $\ln(m_d-m_u)$ may occur due to the
renormalization. However, we  do not
expect that a strong singularity (such as $(m_d-m_u)^{-1}$) that is capable
of eliminating the  original small factor $(m_d-m_u)$ will appear.
In other words, if interaction of a Cooper pair at $k\rightarrow 0$ is
symmetric for protons and neutrons, it is not essential that the
system itself is an asymmetric one, with $n_1\neq n_2$. Such an asymmetry can
be easily adjusted by slightly different Bose chemical potentials, see below,
such that the effective Lagrangian remains symmetric.
Therefore,  when the potential is expressed in terms of Bose chemical
potentials (macrocanonical description) rather than in terms of densities,
we expect the potential retains its  original symmetries.

Therefore, we believe that our approximate $U(2)$ symmetry is
somewhat justified by the original isospin symmetry of the
original protons and neutrons, however this symmetry is not
exactly equivalent to the conventional isotopical $SU(2)$
symmetry. In reality this $U(2)$ symmetry is explicitly slightly
broken, and the potential $V$ has a minimum at $|\psi_1|^2 = n_1,
|\psi_2|^2 = n_2$, where the bulk proton and neutron superfluid
densities $n_1$ and $n_2$ are both non-zero. Hence in the ground
state, $\< |\psi_i|^2 \>= n_i, \, i=1,2$, and both $U(1)$
symmetries are spontaneously broken. In our  letter \cite{typeI},
we presented some general arguments for a generic potential
$V(|\psi_1|^2,|\psi_2|^2)$ which has the symmetry properties
discussed above. In order to be more specific and give the details
of our calculations, in this work we will use the following
standard $\phi^4$-type potential: \bea \label{pot} V &=& - \mu_1
|\psi_1|^2
        - \mu_2 |\psi_2|^2 +\frac{a_{11}}{2} |\psi_1|^4 \nn \\
    &+&  \frac{a_{22}}{2} |\psi_2|^4
    + a_{12} |\psi_1|^2|\psi_2|^2
\eea where $\mu_i$ is the chemical potential of the $i^{th}$
component and $a_{ij}$ is proportional to the scattering length
$l_{ij}$ between the $i^{th}$ and $j^{th}$ components, $a_{ij} = 4
\pi \hbar^2 l_{ij} / \mcoop$. In our nonrelativistic formalism,
the fields $\psi_i$ have energy dimension 3/2 and therefore the
particle density is $n_i = \< |\psi_i|^2 \>$, $i = 1, 2$.

Let's analyze the vacuum structure of our potential (\ref{pot}).
In the limit of exact $U(2)$ symmetry $\mu_i = \mu$, $a_{ij} = a$,
the vacuum manifold is given by the three sphere $|\psi_1|^2
+|\psi_2|^2 = \mu/a$. We are, however, interested in the case when
the $U(2)$ symmetry is explicitly broken to $U(1)\times U(1)$,
giving a particular pattern of proton and neutron condensation
$\<|\psi_i|^2 \>= n_i$. It is natural that even very small $U(2)$
symmetry violating terms are capable of selecting a particular
vacuum on the original degenerate manifold. First, consider the
case when the fourth order couplings are fully degenerate $a_{ij}
= a$, while the chemical potentials are slightly different $\mu_1
= \mu - \delta \mu$, $\mu_2 = \mu + \delta \mu$. In this case, the
condensation pattern is determined solely by the sign of $\delta
\mu$. If $\delta \mu > 0$ then neutrons condense, $\<|\psi_2|^2\>
= \mu_2/a_{22}$, while protons remain uncondensed, $\<\psi_1\>$ =
0; if $\delta \mu < 0$ then protons condense, $\<|\psi_1|^2\> =
\mu_1/a_{11}$, while neutrons remain uncondensed, $\<\psi_2\> =
0$. Observe, that a very small $U(2)$ violating change in chemical
potentials $\mu_1$, $\mu_2$ produces a very large asymmetry of
proton and neutron Cooper pair densities $n_1$, $n_2$.

Now consider a more general situation in which all chemical
potentials $\mu_1, \mu_2$, and fourth order couplings $a_{11},
a_{22}, a_{12}$ are non-degenerate. In this case, a new phase is
possible, where both proton and neutron condensates appear. In
this phase, the proton and neutron Cooper pair densities are given
by: \bea \label{densities}
n_1 &=& \frac{a_{22} \mu_1 - a_{12} \mu_2}{a_{11} a_{22} - a_{12}^2}, \\
\label{densities2} n_2 &=& \frac{a_{11} \mu_2 - a_{12}
\mu_1}{a_{11} a_{22} - a_{12}^2}. \eea This particular vacuum will
be realized if and only if, \bea \label{cond1}
a_{22} \mu_1 - a_{12} \mu_2 > 0,\\
\label{cond2}a_{11} \mu_2 - a_{12} \mu_1 > 0. \eea Notice that
conditions (\ref{cond1},\ref{cond2}) imply $a_{11} a_{22} -
a_{12}^2
> 0$. If conditions (\ref{cond1},\ref{cond2}) are not satisfied then only
one condensate will appear, similar to the case of degenerate
$a_{ij}$'s already described above: if $\mu_1^2/a_{11} <
\mu_2^2/a_{22}$ then $n_2 = \mu_2/a_{22}$, $n_1 = 0$; if
$\mu_1^2/a_{11} > \mu_2^2/a_{22}$ then $n_1 = \mu_1/a_{11}$, $n_2
= 0$.

Throughout the rest of the paper, we will be working in the sector
where both $\psi_1$ and $\psi_2$ obtain non-zero expectation
values, as this is the situation, which is believed to be realized
in neutron star interiors. In this case, the chemical potentials
$\mu_1$, $\mu_2$ are fixed by the equilibrium Cooper pair
densities $n_1$, $n_2$ through equations
(\ref{densities},\ref{densities2}). It is convenient to assume the
following particular parametrization of explicit $U(2)$ violation:
$\mu_1 = \mu -\delta \mu$, $\mu_2 = \mu + \delta \mu$, where
$\delta \mu/\mu \ll 1$, and  $a_{11} = a_{22} = a$, $a_{12} = a -
\delta a$, where $\delta a/a \ll 1$. In this case the equations
for Cooper pair densities (\ref{densities},\ref{densities2})
reduce to: \bea \label{dens_ans1} n_1 &=& \frac{\mu}{2 a - \delta
a} - \frac{\delta \mu}{\delta a} \approx \frac{\mu}{2 a} - \frac{\delta \mu}{\delta a},\\
\label{dens_ans2} n_2 &=& \frac{\mu}{2 a -\delta a} + \frac{\delta
\mu}{\delta a} \approx \frac{\mu}{2 a } + \frac{\delta \mu}{\delta
a} \eea and conditions for stability of the vacuum
(\ref{cond1},\ref{cond2}) reduce to \beq \label{cond_ans}
\frac{|\delta \mu|}{\mu} < \frac{\delta a}{2 a - \delta a} \approx
\frac{\delta a}{2 a} \eeq The approximation made in the second
part of equations (\ref{dens_ans1},\ref{dens_ans2},\ref{cond_ans})
neglects terms of order $\delta a/a$, which are small in the limit
of approximate $U(2)$ symmetry. An important quantity for the
analysis that follows will be the ratio of proton Cooper pair
density to neutron Cooper pair density, $\gamma \equiv n_1 / n_2$.
A typical value of $\gamma$ in the core of a neutron star is
expected to be quite small, $5-15\%$. Thus, we will often use the
limit $\gamma \ll 1$ in our discussion. However, one should remark
here that our qualitative results do not depend on the value of
$\gamma$, as we will explain in what follows. As already
mentioned, the strong deviation of $\gamma$ from $1$ does not
imply a large asymmetry in the interaction between different
species of particles. As is clear from equations
(\ref{dens_ans1},\ref{dens_ans2},\ref{cond_ans}) a value of
$\gamma$ very different from $1$ can be achieved by small $U(2)$
violating terms proportional to $\delta a$, $\delta \mu$ in the
free energy.

The Landau-Ginzburg equations
of motion following from the free energy (\ref{dimfree}) are:
\bea
\label{psi1}
\frac{\hbar^2}{2 \mcoop}(\nabla - \frac{i q}{\hbar c}{\bf
A})^2 \psi_1 = -\mu_1 \psi_1 +  a |\psi_1|^3 \nn \\
    +  (a - \delta a) |\psi_2|^2\psi_1,
\\
\label{psi2}
\frac{\hbar^2}{2 \mcoop} \nabla^2 \psi_2  =
     -\mu_2 \psi_2 + a |\psi_2|^3
    + (a - \delta a) |\psi_1|^2 \psi_2,\\
\label{gauge}
\frac{\nabla\times(\nabla\times{\bf A})}{4 \pi} =
\frac{-i q \hbar}{2 \mcoop c}[\psi_1^{*}(\nabla - \frac{i q}{\hbar c}
{\bf A})\psi_1 - h.c]
\eea
Now let's investigate the structure of proton vortices, which
exist due to the spontaneous breaking of the $U(1)_1$ symmetry.
Such vortices are characterized by the phase of the $\psi_1$ field
varying by an integer multiple of $2 \pi$ as one traverses a
contour around the core of the vortex. By continuity, the field
$\psi_1$ must vanish in the center of the vortex core. Up to this point,
it has been assumed that the neutron order parameter $\psi_2$
will remain at its vacuum expectation value in the vicinity of the
proton vortex. As we have already remarked, this is not the case in many
similar systems.  Actually, it can be shown by using the equations of
motion obtained from the free energy (\ref{dimfree}) that for most
potentials $V$, which are approximately invariant under the
$U(2)$ symmetry, it is {\it impossible} for the $\psi_2$ field to
remain constant when the $\psi_1$ field varies in space. On the
other hand, from the energetic point of view, given that the
potential $V$ can be written approximately as $V \approx
U(|\psi_1|^2 +|\psi_2|^2)$, one can  argue (in part based on
previous work \cite{witten,defectsbook,volovik,BEC,highTc,krstrings,superk})
that it is favorable for
the $\psi_2$ field to increase its magnitude in the vortex core to
compensate the decrease in the magnitude of $\psi_1$.

So, anticipating a non-trivial behavior of the neutron field
$\psi_2$, let's adopt the following cylindrically symmetric ansatz
for the fields describing a proton vortex with a unit winding
number:
\bea
\label{Ansatz}
\psi_1 &=& \sqrt{n_1}~f(r)~e^{i \theta}, \nn \\
\, \psi_2 &=& \sqrt{n_2}~g(r), \\
{\bf A} &=& \frac {\hbar c}{q}
\frac{a(r)}{r} ~\hat{\theta} \nn
\eea
where $(r,\theta)$ are the
standard polar coordinates. Here we assume that the proton vortex
is sufficiently far from any rotational neutron vortices, so that
any variation of $\psi_2$ is solely due to the proton vortex. The
functions $f$, $g$, and $a$ obey the following boundary
conditions: $f(0) = 0$, $f(\infty) =1$, $g'(0) = 0$, $g(\infty) =
1$, $a(0) = 0$, and $a(\infty) =1$. We see that the fields
$\psi_1$ and $\psi_2$ approach their vacuum expectation values at
$r = \infty$.

The London penetration depth $\lambda$ and the coherence length
$\xi$ of the proton superconductor will be introduced in the
standard fashion:
\bea
\label{GLparams}
\lambda &=& \sqrt{\frac{\mcoop c^2}{4 \pi q^2 n_1}}, \\
\xi &=& \sqrt{\frac{\hbar^2}{2 \mcoop n_1 a}}.
\eea
We wish to find the asymptotic behavior of fields $\psi_1$,
$\psi_2$ and ${\bf A}$ far from the proton vortex core, as this
will determine whether distant vortices repel or attract each
other. The asymptotic behavior can be found analytically by
expanding the fields defined in (\ref{Ansatz}):
\bea
\label{asympsoln}
f(r) &=& 1 + F(r), \nn \\
g(r) &=& 1 + G(r), \\
a(r) &=& 1 - r S(r), \nn
\eea
so that far
away from the vortex core, $F, G, r S \ll 1$ and $F, G, S
\rightarrow 0$ as $r \rightarrow \infty$. This allows us to
linearize far from the vortex core the equations of motion
(\ref{psi1},\ref{psi2},\ref{gauge})
corresponding to the free energy (\ref{dimfree}) to obtain:
\bea
\label{FG}
(\frac{\d^2}{\d r^2} + \frac{1}{r}\frac{\d}{\d r}) \bmat F \\
G \emat &=& {\bf M} \bmat F \\ G \emat, \\
\label{S}
S'' + \frac{1}{r} S' - \frac{1}{r^2} S &=& \frac{1}{\lambda^2} S,
\eea
where all derivatives are with respect to $r$ and the
matrix ${\bf M}$ mixing the fields $F$ and $G$ is,
\beq
\label{mixmat}
{\bf M} = \frac{4 \mcoop}{\hbar^2} \bmat a & a - \delta a \\
a - \delta a & a \emat \bmat n_1 & 0 \\
0 & n_2 \emat
\eeq
Here we assume that $(r S)^2 \ll F, G$, i.e. the superconductor is
not in the strong type-II regime (this is justified since we are
only attempting to find the boundary between type-I and type-II
superconductivity). The solution to Eq. (\ref{S}) is known to be:
\beq \label{Ssol} S = \frac{C_A}{\lambda} K_1(r/\lambda) \eeq
where $K_1$ is the modified Bessel function and $C_A$ is an
arbitrary constant. The remaining equation (\ref{FG}) can be
solved by diagonalizing the mixing matrix ${\bf M}$. In previous
works the influence of the neutron condensate on the proton vortex
was neglected, which formally amounts to setting the off-diagonal
term $M_{12}$ in Eq.~(\ref{mixmat}) to $0$. In that case, one can
assume that the neutron field remains at its vacuum expectation
value, i.e. $G = 0$, to obtain, \beq F = C_F K_0(\sqrt{2} r/\xi)
\eeq where $K_0$ is the modified Bessel function. It is estimated
that $\lambda \sim 80$~fm and $\xi \sim 30$~fm \cite{link}, which leads to
$\kappa = \lambda / \xi \sim 3$ for the Landau-Ginzburg parameter.
As is known from conventional superconductors, if $\kappa >
1/\sqrt{2}$, distant vortices repel each other leading to type-II
behavior. This is the standard picture of the proton
superconductor in neutron stars that is widely accepted in the
astrophysics community.

However, the standard procedure described above is inherently
flawed since the system exhibits an approximate $U(2)$ symmetry,
and therefore the couplings $a_{ij}$ are approximately equal
$a_{11} \approx a_{22} \approx a_{12}$. This makes the mixing
matrix ${\bf M}$ nearly degenerate. The general solution to Eq.
(\ref{FG}) is:
\beq \label{soln} \bmat F \\ G \emat = \sum_{i=1,2}
C_i K_0(\sqrt{\nu_i} r) ~{\bf v_i}
\eeq
where $\nu_i$ and ${\bf
v_i}$ are the eigenvalues and eigenvectors of matrix ${\bf M}$,
and $ C_i$ are constants to be calculated by matching to the
solution of the original nonlinear equations of motion.
We would like to introduce the parameter $\epsilon$
(which measures the asymmetry between the proton and neutron Cooper pairs)
defined in the following way:
\beq
\epsilon= (a_{11} a_{22} -a_{12}^2)/a_{ij}^2 \simeq 2 \frac{\delta a}{a}.
\eeq
We should remark here that our qualitative results do not depend on
the value of $\gamma \equiv n_1/n_2.$
Indeed, no matter what $n_1$ and $n_2$ are, the mixing
matrix is still singular in the limit $\epsilon \rightarrow 0$.
Hence, we still get one eigenvalue which vanishes when
$\epsilon \rightarrow 0$. So, the only crucial assumption
is $\epsilon \ll 1$. However, to simplify our formula for the eigenvalues
in what follows we assume a specific value of $\gamma \equiv n_1 / n_2 \ll 1$.
It simply allows our results to be expressed in a more transparent way.
In the limit $\gamma = n_1/n_2 \ll 1$ and $\epsilon = 2 \delta a /a \ll 1$
one can estimate the eigenvalues and eigenvectors of the matrix ${\bf M}$ as:
\bea
\label{eigen1}
\nu_1 &\simeq& \frac{2 \epsilon}{\xi^2},~~~~
{\bf v_1} \simeq \bmat -1 \\ \gamma \emat,\\
\label{eigen2}
\nu_2 &\simeq& \frac{2}{\gamma \xi^2},~~
{\bf v_2} \simeq \bmat 1 \\ 1 \emat.
\eea
The physical meaning of solution (\ref{soln}) is simple: there are
two modes in our two component system. The first mode describes
fluctuations of relative density (concentration) of two components
and the second mode describes fluctuations of overall density of
two components. Notice that $\nu_1 \ll \nu_2$, and hence the
overall density mode has a much smaller correlation length than
the concentration mode. Therefore, far from the vortex core, the
contribution of the overall density mode can be neglected, and one
can write:
 \beq
 \label{assymFG}
\bmat F \\ G \emat (r\rightarrow\infty)
 \simeq C_1 K_0(\sqrt{2 \epsilon} r/\xi) \cdot \bmat -1 \\
 \gamma \emat
\eeq
The most important result of the above discussion is that the
distance scale over which the proton and neutron condensates tend
to their vacuum expectation values near a proton vortex is of
order $\xi/\sqrt{\epsilon}$ - the correlation length of the
concentration mode. Since $\epsilon \ll 1$, this distance scale
can be much larger than the proton correlation length $\xi$, which
is typically assumed to be the radius of the proton vortex core.
The appearance of the concentration mode is not surprising since
we presented some arguments (before calculations)
 supporting the picture that the neutron condensate will increase its
magnitude slightly in the vortex core, while the proton condensate
will decrease its magnitude to $0$ in the core center. We note
that in the limit $\epsilon \rightarrow 0$ the size of the proton vortex
core becomes infinitely large, and the vortex is thereby destroyed. This
is in accordance with the topological arguments, which state that if
the $U(2)$ symmetry were exact with
 $\epsilon \equiv 0$, and it is spontaneously broken to $U(1)$,
there will be  no stable vortices in the system.

We have also verified the above results numerically by
solving the equations of motions
(\ref{psi1},\ref{psi2},\ref{gauge})
with a particular choice of the approximately $U(2)$ symmetric
interaction potential $V$.
 Our numerical
results support the analytical calculations given above. Namely,
we find that the magnitude of the neutron condensate is slightly
increased in the vortex core, the radius of the magnetic flux tube
is of order $\lambda$ and the radius of the proton vortex core is
of order $~\xi/\sqrt{\epsilon}$.
\begin{figure}[t]
 \begin{center}
\psfrag{y}{\large{$f,G/\gamma,a$}} \psfrag{x}{\large{$~~~~\rtil$}}
\includegraphics[width=0.45\textwidth]{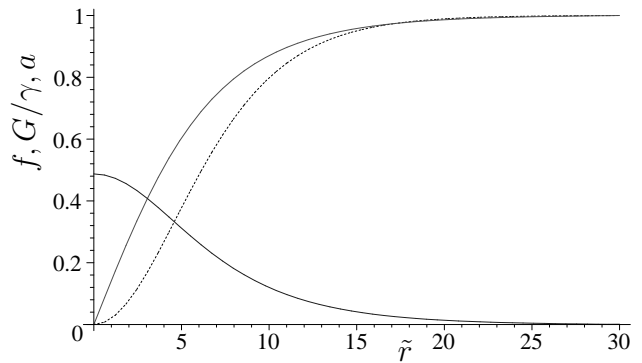}
\caption{ In this figure we show the functions $f(\rtil)$,
$G(\rtil)/\gamma$, and $a(\rtil)$ (defined in Eqs.
(\ref{Ansatz},\ref{asympsoln})) as a function of the dimensionless
radial coordinate $\rtil= r/\xi$. The dotted line corresponds to
$a(\rtil)$, the solid line approaching $1$ at large $\rtil$
corresponds to $f(\rtil)$, and the solid line approaching $0$ at
large $\rtil$ corresponds to $G(\rtil)/\gamma$.
\label{fig:npvortex}}
 \end{center}
\end{figure}
In Fig. \ref{fig:npvortex} we show the numerical solution of
the profiles of the proton vortex ($f(\rtil)$),
neutron condensate ($G(\rtil)/\gamma$), and $a(\rtil)$
(related to the gauge field through Eq. (\ref{asympsoln})) as a
function of the dimensionless radial coordinate $\rtil = r/\xi$,
where $\xi$ is the coherence length (\ref{GLparams}).
We have used
$\kappa = 3$, $n_1/n_2 = 0.05$, and $\epsilon = 0.02$ in this
numerical solution.

\section{Vortex-vortex interaction}

Now that we know the approximate solution for the proton vortex,
we will proceed to look at the interaction between two widely separated
proton vortices. If the interaction between two
vortices is repulsive, it is energetically favorable for the
superconductor to organize an Abrikosov vortex lattice with each
vortex carrying a single magnetic flux quantum. As the magnetic
field is increased, more vortices will appear in the material.
This is classic type-II behavior. If the interaction between two
vortices is attractive, it is energetically favorable for $n$
vortices to coalesce and form a vortex of winding number $n$.
This is type-I behavior.
Typically, the Landau-Ginzburg parameter $\kappa = \lambda/ \xi$
is introduced. In a conventional superconductor, if $\kappa <
1/\sqrt{2}$ then the superconductor is type-I and vortices
attract. If $\kappa>1/ \sqrt{2}$ then vortices repel each other and
the superconductor is type-II. As mentioned above, the typical value
for a neutron star is $\kappa \sim 3$, so one could naively expect
that the proton superfluid is a type-II superconductor.

The case we considered in the previous section has
one new element, $\epsilon$ which was not present in the standard
type-I/II classification. However,
we expect that analogous classification should remain in effect
for the proton vortices described above.
In such an analysis the coherence
length $\xi$ should be replaced by the actual
size of the proton vortices ${\delta} \sim \xi/\sqrt{\epsilon}$.
Therefore, we will define a new Landau-Ginzburg parameter for our
case,
\beq
\label{kappaeff}
\kappa_{np} = \frac{\lambda}{\delta} = \sqrt{\epsilon} \frac{\lambda}{\xi}.
\eeq
We expect type-I behavior
with vortices attracting each other if $\kappa_{np} \ll 1$ and
type-II behavior if $\kappa_{np} \gg 1$. For relatively small
$\epsilon$ such an argument would immediately suggest that for the
typical parameters of the neutron stars type-I  superconductivity is
realized (rather than the naively assumed type-II superconductivity).
In what follows we present several different arguments supporting
this claim.

\subsection{Calculating in the intervortex potential: method I}

In order to make this qualitative discussion more concrete, we
will present three different calculations supporting our claim
that for the typical parameters of a neutron star
the proton superconductor may be type-I rather than type-II.
First of all, we follow the method suggested originally in
\cite{speight} to calculate the force between two widely separated
vortices. The methods of
\cite{speight} were subsequently applied in \cite{mack} to the
case similar to ours, the interaction of two widely separated
vortices that have nontrivial core structure. In these
papers, the force between two widely separated vortices is
calculated by using a linearized theory with point sources added
at the location of the vortices. The point sources in the
linearized theory are chosen to produce fields matching the long
distance asymptotics of the original theory. Therefore, we expand
the free energy (\ref{dimfree}) up to quadratic order in the
fields $F$, $G$, and ${\bf A}$ introduced in the previous section
eliminating the
phase of $\psi_1$ field in favor of the longitudinal component of
{\bf A}, to produce a non-interacting free energy ${\cal F}_{free}$.
\bea
\label{Free}
{\cal F}_{free} &=& \int d^2 r
  \bigg[\frac{\hbar^2}{2 \mcoop}(n_1 (\nabla F)^2
  + n_2 (\nabla G)^2)  \nn \\
  &+&  \frac{1}{8 \pi}((\nabla \times {\bf A})^2 + \frac{1}{\lambda^2}
  {\bf A}^2) + 2 a n_1^2 F^2 \nn \\
  &+& 4 (a - \delta a) n_1 n_2  F G + 2 a n_2^2 G^2 \bigg]
\eea
Following \cite{speight,mack}, we must also add the source terms
for each field to model the vortices:
\beq
\label{freesource}
{\cal F}_{source} = \int d^2 r
    ( \rho F + \tau G + {\bf j} \cdot {\bf A}).
\eeq
The solutions to the equations of motion following from
${\cal F}_{free}$ coupled to the sources can be obtained in the
same manner as was done in \cite{mack}. The equations of motion
resulting from ${\cal F}_{free} + {\cal F}_{source}$ are:
\bea
\label{sourceeomFG}
(\nabla^2 - {\bf M}) \bmat F \\ G \emat &=& \frac{\mcoop}{\hbar^2}
    \bmat \rho/n_1 \\ \tau/n_2 \emat, \\
\label{sourceeomA}
\nabla^2 {\bf A} - \frac{1}{\lambda^2} {\bf A} &=& 4 \pi {\bf j}
\eea
where ${\bf M}$ is the same mixing matrix given in Eq.~(\ref{mixmat}).
Since we are interested in the asymptotic behavior, we will choose the
the first eigenvalue/eigenvector solution (\ref{eigen1}) that
diagonalized ${\bf M}$. Therefore, the asymptotic field solutions are
given by Eqs. (\ref{Ssol},\ref{assymFG}):
\bea
\label{solnfields}
F \simeq -G/\gamma &\simeq& - C_1 K_0(\sqrt{2 \epsilon} r/\xi), \\
{\bf A} &\simeq& \frac{\hbar c}{q \lambda}C_A K_1(r/\lambda)~\hat{\theta}. \nn
\eea
Following \cite{mack}, we require that (\ref{solnfields}) are solution
of the equations of motion (\ref{sourceeomFG}) and (\ref{sourceeomA}).
The source solutions can immediately be written down when we compare Eqs.
(\ref{sourceeomFG}) and (\ref{sourceeomA}) with the following Bessel
equations:
\bea
(\nabla^2 - \mu^2) K_0(\mu x) &=& - 2 \pi ~\delta({\bf x}) \\
(\nabla^2 - \mu^2) \frac{x_j}{x} K_1(\mu x) &=&
    \frac{2 \pi}{\mu} ~\d_j \delta({\bf x})
\eea
The source solutions that solve the equations of motion along
with (\ref{solnfields}) are:
\bea
\label{solnsource}
\rho \simeq -\tau &\simeq& \frac{1}{2}{\Big(\frac{\hbar c}{q
\lambda}\Big)}^2 C_1 \delta^2({\bf r}), \\
{\bf j} &\simeq& -\frac{\hbar c}{2 q} C_A
\nabla \times(\delta^2({\bf r}) \hat{z}) \nn
\eea
Since the equations of motion corresponding to
${\cal F}_{free}$ coupled to ${\cal F}_{source}$ are linear,
the two-vortex
solution can be considered as the sum of two single vortices at
positions ${\bf r}_1$ and ${\bf r}_2$.
To calculate the vortex-vortex  interaction energy, we use ansatz
$(F,G,{\bf A}) = (F_1+ F_2, G_1+G_2,{\bf A_1}+{\bf A_2})$
$(\rho,\tau,{\bf j}) =
(\rho_1+\rho_2,\tau_1+\tau_2,{\bf j_1}+{\bf j_2})$
in ${\cal F}_{free} + {\cal F}_{source}$
and subtract off the energy of each isolated vortex.
The notation $1,2$ indicates that these are functions of
${\bf r} - {\bf r}_{1,2}$. Using the equations of motion
(\ref{sourceeomFG},\ref{sourceeomA}) the interaction energy
can be written as
\beq
\label{fint}
{\cal F}_{int} = \int d^2 r ( {\bf j_1} \cdot {\bf A_2} + \rho_1 F_2
    + \tau_1 G_2  )
\eeq
Substituting the asymptotic field solutions
(\ref{solnfields},\ref{solnsource}) into the interaction energy given above,
the integration can be done as in \cite{mack} to obtain the
following expression for the interaction energy per unit vortex
length of two widely separated parallel vortices:
\bea
\label{potential}
U(d) &\simeq& \frac{2 \pi \hbar^2
n_1}{\mcoop}(C_A^2 K_0(d/\lambda) \nn \\
    &-& C_1^2 (1 + {\cal O}(\gamma)) K_0(\sqrt{2 \epsilon}d/\xi))
\eea
where $d = | {\bf r}_1 - {\bf r}_2 | \rightarrow \infty$
is the separation between the two vortices. We see that
if the first term in $U$ dominates as $d \rightarrow \infty$ then
the potential is repulsive, otherwise, if the second term
dominates the potential is attractive. In other words, if
$\sqrt{\epsilon}\lambda/\xi < 1/\sqrt{2}$, then vortices attract
each other and the superconductor is type-I; otherwise, vortices
repel each other and the superconductor is type-II. This confirms
our original qualitative argument that $\kappa_{np} =
\lambda/\delta = \sqrt{\epsilon} \lambda/\xi$ should be considered
as an effective Landau-Ginzburg parameter, which determines the
boundary between the type-I and type-II proton superconductivity.
In terms of the parameters of our theory, we have
\beq
\label{kappa1}
\kappa_{np} =
\frac{\lambda}{\delta} = \sqrt{\epsilon} ~\frac{\lambda}{\xi}=
\frac{\mcoop c}{\sqrt{\pi} \hbar} \frac{\sqrt{\da}}{q}.
\eeq
In this case we see that the type-I/II behavior is controlled in part
by the degree of symmetry breaking (proportional to $\da$).
Our numerical estimates (see conclusion) suggest
that $\kappa_{np} <1/\sqrt{2}$, and therefore, the system is
a type-I superconductor.

\subsection{Calculating the intervortex potential: method II}

Due to the importance and far reaching consequences of this
result, we have also calculated the vortex-vortex  interaction
energy in a more direct way following \cite{Kramer},
where the introduction of the auxiliary sources is
completely avoided. This method, very different
in nature, has produced the same result (\ref{potential}) as the
above procedure, therefore confirming our picture.

We proceed as follows. First, we again eliminate the phase of
$\psi_1$ with an appropriate gauge transformation. Now, recalling
our definitions (\ref{Ansatz}, \ref{asympsoln}), let $(F, G, {\bf
A}) = (F_1, G_1, {\bf A_1})$ be the exact fields of a single
vortex located at $r_1$. Also let $(F, G, {\bf A}) = (F_1+F_2, G_1
+ G_2, {\bf A_1} + {\bf A_2})$ be the exact fields produced by two
vortices at locations ${\bf r_1}$ and ${\bf r_2}$. Note that
subscript $2$ does not generally refer here to functions of ${\bf
r} - {\bf r_2}$ as it did in the previous subsection: here $(F_2,
G_2, {\bf A_2})$ are just corrections to the fields of the single
vortex at ${\bf r_1}$. When ${\bf r}$ is far from the core of
vortex $1$, $F_1, G_1, {\bf A_1}$ are small and the equations of
motion can be linearized to yield familiar asymptotics
(\ref{solnfields}). Moreover, when ${\bf r}$ is far from the cores
of both vortices,  $F_i, G_i, {\bf A_i}$ are small for both $i =
1,2$ and we obtain:
\bea
\label{solnfields2}
F_i \simeq -G_i/\gamma &\simeq& - C_1 K_0(\sqrt{2 \epsilon} |{\bf r} - {\bf
r_i}|/\xi),  \\
{\bf A_i} &\simeq& \frac{\hbar c}{q \lambda}C_A
K_1(|{\bf r} - {\bf r_i}|/\lambda)~\hat{\theta}. \nn
\eea
To calculate
the interaction energy of the two vortices, let us divide the
space into two cells $T_1$ and $T_2$, which contain the centers of
vortices 1 and 2 respectively (if we had more than two vortices in
our problem these would be the Wigner-Seitz cells of the vortex
lattice). For simplicity, we will take the boundary between two
cells to be the line perpendicular to and passing through the
middle of the line-segment joining the cores of two vortices. This
boundary then acts as an axis of symmetry for the problem: the
total energy of two interacting vortices is just twice the energy
of each cell. The vortex-vortex interaction energy is then:
\bea
\label{int2}
{\cal F}_{int} &=&{\cal F}[F_1 + F_2, G_1 + G_2,
{\bf A_1} + {\bf A_2}] -2 {\cal F}[F_1, G_1, {\bf A_1}] \nn \\
&=& 2 \int_{T_1} d^2 r ({\cal E}[F_1+F_2, G_1 + G_2, {\bf A_1} +
{\bf A_2}] \nn \\
&-& {\cal E}[F_1, G_1, {\bf A_1}]) - 2\int_{T_2} d^2 r
{\cal E}[F_1, G_1, {\bf A_1}]
\eea
where ${\cal E} = d{\cal F}/dr^2$ is the free energy per
unit volume and subscripts $T_i$ indicate integration over the
corresponding cells. We assume the vortex separation to be
large, therefore the fields $F_1, G_1, {\bf A_1}$ inside cell $T_2$ will be
small, so we can expand the integral over the cell $T_2$ in
(\ref{int2}) to second order in $F_1, G_1, {\bf A_1}$. Similarly,
the corrections to fields $F_1, G_1, {\bf A_1}$ inside cell $T_1$
will be small so we can expand the integral over the cell $T_1$ in
(\ref{int2}) to second order in corrections $F_2, G_2, {\bf A_2}$.
The resulting expression can be simplified by using equations of
motion (\ref{psi1}), (\ref{psi2}), (\ref{gauge}) to obtain:
\begin{widetext}
\bea
\label{intsurf}
{\cal F}_{int} &\simeq& 2 \oint_{T_1} {\bf dS}
\cdot (\frac{\hbar^2 n_1}{2 \mcoop} F_2 (2 \nabla F_1 + \nabla F_2)
+ \frac{\hbar^2 n_2}{2 \mcoop}G_2(2 \nabla G_1 + \nabla G_2)
+ \frac{1}{8 \pi}{\bf A_2}\times(2 \nabla \times {\bf A_1}
+ \nabla \times {\bf A_2})) \nn\\
&-& 2\oint_{T_2} {\bf dS}\cdot(\frac{\hbar^2 n_1}{2
\mcoop}F_1 \nabla F_1 + \frac{\hbar^2 n_2}{2 \mcoop} G_1 \nabla G_1 +
\frac{1}{8 \pi} {\bf A_1}\times(\nabla \times {\bf A_1}))
\eea
\end{widetext}
Here the integrals are over the boundary of cells $T_1$ and $T_2$.
Since this boundary is far away from either vortex center, we can
use the asymptotic expressions (\ref{solnfields2}) for the fields $(F_i,
G_i, {\bf A_i})$ to explicitly calculate the surface integrals in
(\ref{intsurf}). We note that by symmetry of the asymptotic
solution on the boundary, the second integral in (\ref{intsurf})
cancels with the part of the first integral to yield:
\bea
\label{intsurf2}
{\cal F}_{int} &\simeq& 2 \oint_{T_1} {\bf dS}
\cdot (\frac{\hbar^2 n_1}{\mcoop} F_2 \nabla F_1 + \frac{\hbar^2
n_2}{\mcoop} G_2  \nabla G_1  \nn \\
&+& \frac{1}{4 \pi}{\bf A_2}\times(\nabla \times {\bf A_1}))
\eea
Substituting asymptotic solutions
(\ref{solnfields2}) into the above, we find the vortex-vortex
interaction energy per unit length to be:
\bea
U(d) &=& \frac{2 \pi
\hbar^2 n_1}{\mcoop}(C_A^2 K_0(d/\lambda)  \nn \\
&-& C_1^2 (1+ {\cal O}(\gamma)) K_0(\sqrt{2 \epsilon} d/\xi))
\eea
in accordance with our previous
result (\ref{potential}).

\section{Critical Magnetic Fields}

Our third check of the main result that for relatively
small $\epsilon$ the superconductor in the neutron stars may be,
in fact, type-I is based on the calculation of the
critical magnetic fields.
Usually one calculates the critical  magnetic fields $H_c$ and
$H_{c2}$. These are the physically meaningful fields above which
the superconductivity is destroyed in type-I and type-II
superconductors respectively. If $H_c> H_{c2}$ then the
superconductor is type-I, otherwise, the superconductor is type-II.

First, we will calculate the critical magnetic
field $H_c$. This is defined as the point at which the Gibbs free
energy of the normal phase is equal to the Gibbs free energy of
the superconducting phase. In other words, as the external magnetic
field $H$ is increased above $H_c$, it is energetically favorable
for the superconducting state to be destroyed macroscopically.
The Gibbs free energy in the presence
of an external magnetic field $H$ is:
\beq
g(H,T) = f(B,T) - \frac{B H}{4 \pi}
\eeq
where $H$ is the external magnetic field,  $B$ is the magnetic
induction, and $T$ denotes temperature.
The quantity $f(B,T)$ is the integrand of the free energy density (given by
Eq.~(\ref{dimfree})) over a superconducting
sample. For the superconducting state where
$\pvev = n_1$, $\nvev = n_2$, and $B=0$ (Meissner effect),
the Gibbs free energy is
\beq
g_s(H,T) = - \frac{\mu^2}{2a}  -\frac{(\delta\mu)^2}{\delta a}
-\delta a \left( \frac{\mu}{2a} \right)^2,
\eeq
where we expressed the result in terms of the
parameters of Eq. (\ref{dimfree}).
We also replaced the densities $(n_1+n_2)\rightarrow \mu/ a$ and
$(n_2-n_1)\rightarrow 2 \delta\mu / \delta a$ in terms of the same
parameters by neglecting small factors $\sim \delta a/a$.
For the normal state, we have
$\pvev = 0$, $\nvev = (\mu + \delta \mu)/a$, and
$B=H$. The Gibbs free energy is:
\beq
g_n(H,T) = - \frac{H^2}{8 \pi} - \frac{\mu^2}{2a}
    -\frac{\mu\delta\mu}{a}
\eeq
As $H_c$ is defined as the point at which $g_s(H_c) = g_n(H_c)$,
we can solve for the critical field.
Equating the free energies of the normal and superconducting
state, we find
\beq
\label{hc}
H_c = \sqrt{8 \pi   \delta a   }\left(\frac{\mu}{2a}-
\frac{\delta\mu}{\delta a}\right)
\rightarrow n_1\sqrt{8 \pi   \delta a   } ,
\eeq
where at the final stage we used the equation for $n_1$
in the superconducting phase expressed
in terms of the original parameters (\ref{densities}).

Now we will proceed to calculate $H_{c2}$. This is the
critical magnetic field below which it becomes energetically
favorable for a microscopic region
of the superconducting state to be nucleated, with the normal state
occurring everywhere else in space. In order to calculate
$H_{c2}$, we follow the standard procedure and linearize the equations
of motion for $\psi_1$ about the normal state with $\pvev =0$ and
$\nvev = (\mu + \delta \mu)/a$.
The linearized equation of motion reads,
\bea
\label{landaulevel}
\frac{\hbar^2}{2 \mcoop} \left(-i \nabla
    - \frac{q}{\hbar c} {\bf A} \right)^2 \psi_1 = \omega \psi_1, \\
\label{landauenergy}
\omega = (\mu + \delta \mu) \frac{\da}{a} -  2 \delta \mu.
\eea
This is simply a Schrodinger equation for a particle in a
magnetic field, with an energy of $\omega$. This is a standard quantum
mechanics problem and we can immediately write down the solution. The
first Landau level is the ground state energy of
$\epsilon_0(H) = \hbar |q| H/2 \mcoop c$. Therefore,
if $\omega <\epsilon_0$, then only the
trivial solution with $\psi_1 = 0$ is possible. The critical
field $H_{c2}$ is defined as the point at which
$\omega = \epsilon_0(H_{c2})$. This is given as
\beq
\label{hc2}
H_{c2} = \frac{2 \mcoop c}{\hbar |q|}
    [(\mu + \delta \mu) \frac{\da}{a} - 2 \delta \mu]
\simeq \frac{4 \mcoop c}{\hbar |q|}\da~ n_1
\eeq
Now that we have the critical fields $H_c$ and $H_{c2}$ in hand, we
can compare the two in order to determine the type-I/II nature. If
$H_c < H_{c2}$ this means that it is energetically
favorable for microscopic regions of the superconducting state
to be nucleated as $H$ is decreased. This is type-II behavior, and
this nucleation manifests itself in the form of an
vortex lattice. If $H_c > H_{c2}$, then it is energetically
favorable for macroscopic regions of the superconducting state
to be present as $H$ is decreased. This is a type-I superconductor and
the superconducting state persists
everywhere in the material when $H < H_c$, as opposed to a type-II
superconductor where it is localized in regions of space in between
the vortices. From
Eqs.~(\ref{densities},\ref{hc},\ref{landauenergy},\ref{hc2}) we see that
\beq
\label{ratioH}
\frac{H_{c2}}{H_c} \simeq \sqrt{2} \frac{\mcoop c}{\sqrt{\pi} \hbar}
    \frac{\sqrt{\da}}{q} = \sqrt{2} ~ \kappa_{np}.
\eeq
This agrees
with the parametrical behavior given in Eq. (\ref{kappa1}) obtained
from the vortex interaction calculation of the previous section.
To estimate $H_c$ numerically, it is convenient to represent $H_c$ as
\beq
H_c = \frac{\varphi_0}{2 \pi \lambda \xi} \sqrt{\frac{\delta a}{a}},~~
\varphi_0 = \frac{2 \pi \hbar c}{q}
= 2 \times 10^7 {\mathrm{G}} \cdot {\mathrm{cm}}^2,
\eeq
where $\varphi_0$ is the quantum of the fundamental flux.
If we substitute $\lambda = 80$~fm and $\xi=30$~fm (typical values)
in the expression for the critical magnetic field (\ref{hc}),
$H_c$ is estimated to be the $H_c \simeq 10^{14}$~G, which is smaller
than the ``naive'' estimate by a factor of $\sqrt{\delta a/ a} \sim 10^{-1}$.
It is quite amazing that very different calculations
of the critical magnetic fields (\ref{ratioH}) lead exactly to the
same conclusion which was derived from analysis of the vortex-vortex
interaction (\ref{kappa1}).

\section{Conclusion}

In this paper we have demonstrated using various calculations that
the proton superfluid present inside a neutron star may in fact be a
type-I superconductor \cite{typeI}. This supports the observation
made my Link \cite{link} that the conventional picture of
type-II superconductivity may be inconsistent with the
observations of long period precession in isolated pulsars \cite{precession}.
The most important consequence of this paper is that whether the
proton superconductor is type-I or type-II depends strongly on the
magnitude of the $SU(2)$ asymmetry parameter $\epsilon$.
Specifically, we find that the superconductor is type-I when
$\kappa_{np} = \sqrt{\epsilon} \lambda/\xi < 1/\sqrt{2}$, and
type-II otherwise. This result is quite generic, and not very
sensitive to the specific details  of the interaction potential $V$.
In particular, when $\epsilon \rightarrow 0$ the superconductor is
type-I. The parameter $\epsilon$ is not known precisely; the
corresponding  microscopical calculation would require the
analysis of the scattering lengths of Cooper pairs for different
species. We can roughly estimate this parameter  as being
related to the original $SU(2)$ isospin symmetry breaking
$\epsilon \sim (m_n - m_p)/m_n \sim 10^{-2}$. If we assume a typical
value for $\lambda/\xi \sim 3$ and $\epsilon \sim 10^{-2}$,
we estimate $\kappa_{np}
=\sqrt{\epsilon} \lambda/\xi \sim 0.3 < 1/\sqrt{2}$, which
corresponds to a type-I superconductor. From these crude
estimates, we see that it is very likely that neutron stars are
type-I superconductors with the superconducting region devoid of
any magnetic flux, as was originally suggested in \cite{link} to
resolve the inconsistency with observations of long period
precession \cite{precession} in isolated pulsars.
Many mechanisms have been proposed to explain glitches
\cite{glitch1}.
If the proton superfluid exhibits type-I
superconductivity then some explanations of glitches \cite{glitchflux}
that assume type-II superconductivity would have to be reconsidered,
as suggested in Ref. \cite{link}.
It might be interesting to consider how the presence of a nonzero
proton condensate affects the characteristics of the  neutron
vortices that carry angular momentum. In particular, as we have demonstrated
for magnetic flux tubes in this paper, the neutron vortices might
have an enhanced proton superfluid density inside, as well as a
coherence length $\xi_n$ (the approximate size of the vortex)
which is much larger than originally expected.
In this case, the pinning force that is related to the size of the
vortex core \cite{pinning} could be very different from the
simple estimates when the strong interaction between
the neutron and proton Cooper pairs is ignored.

If the core is indeed a type-I superconductor, the magnetic
field could exist in macroscopically large regions where there are
alternating domains of superconducting (type-I) matter and normal matter.
In this case, neutron stars could have long period precession.
Such a structure follows from few different arguments.
First of all, as has been
estimated \cite{Baym}, it takes a very long time   to
expel a typical magnetic flux from the neutron star core.
Therefore, if the magnetic field existed before the neutron star became
a type-I superconductor, it is likely that magnetic field will remain there.
Another argument suggesting the same outcome follows from the fact
that topology (magnetic helicity) is frozen in the environment
with high conductivity; therefore, the magnetic field must remain in
the bulk of the neutron star. The last argument supporting the same picture
is due to Landau \cite{LandauLam} who argued that if a body of arbitrary shape
(being a type-I superconductor) is placed under influence of the
external magnetic field with strength $H < H_c$, then   the magnetic field
in  some parts of the body may reach the critical value $H_c$,
while in other parts of the body it may remain smaller than the critical
value, $H < H_c$. In this case, a domain structure will  be formed, similar
to ferromagnetic systems.  Specifically, on a macroscopic
distance scale, the magnetic flux must be embedded in the superconductor.
This would mean that the superconductor is in an intermediate
state as opposed to the vortex state of the type-II superconductor, which was
assumed to be realized up to this point.
The intermediate state is characterized by
alternating domains of superconducting and normal
matter. The superconducting domains will then exhibit the Meissner effect,
while the normal domains will carry the required magnetic flux. The pattern
of these domains is usually strongly related to the
geometry of the problem. The simplest geometry, originally considered by
Landau \cite{LandauLam}, is a laminar structure of alternating superconducting
and normal layers. However, it has been also argued \cite{Sedrakian} that due
to the geometry of the neutron vortex lattice, the
normal proton domains will be in the shape of cylinders concentric
with the rotational neutron vortices.

While some precise calculations are required
for understanding of the magnetic structure in this case, one can
give some simple  estimation of  the size of the domains
using the calculations Landau presented for a different geometry.
His formula \cite{LandauLam} suggests that the typical size of a
domain is
\beq
\label{domain}
a\sim 10 \sqrt{R \Delta},
\eeq
where $R$ is a typical external size identified with a neutron star core
($R\sim 10$~km), while $\Delta$ is a typical width of the domain wall
separating normal and superconducting states. We estimate
$\Delta\sim \delta =\xi/\sqrt{\epsilon}$ as the largest microscopical
scale of the problem. Numerically, $a\sim 10^{-1}$~cm which implies that a
typical domain with size of order $\sim 10^{-1}$~cm can accommodate
$\sim 10 $ neutron vortices separated by a distance $\sim 10^{-2}$~cm.

The consequences of this picture still remain to be explored.
In particular, it might be of importance in the explanation of
glitches. It may be also important in analysis of the cooling
properties of the neutron stars.

It would be very interesting to test the ideas outlined in this paper
by doing laboratory experiment in the spirit of the
Cosmology in the Laboratory (COSLAB) program.
In particular, it would be interesting to find a condensed matter
system (high $T_c$ superconductor?) where the very interesting
feature discussed in this paper can be tested.
Namely, the core of the vortices
and their interactions could be very different from the simple estimates.
This might happen if there is a condensate of another field
with energy scales almost degenerate with energies determined by
the  Landau-Ginsburg complex $\psi$ field describing a superconductor.
Over the last few years
several experiments have been done to test ideas drawn from
cosmology and astrophysics
(see Ref. \cite{volovik} and the web page \cite{COSLAB} of the latest
COSLAB meeting for further details).

\section*{Acknowledgments}

We are grateful to M. Prakash, M. Alford, and J. Lattimer for
bringing reference \cite{link} to our attention. We are thankful to
B. Link for his useful remarks. We would
also like to thank D. T. Son and M. A. Stephanov for discussions and
their useful remarks. This work was
supported in part by the Natural Sciences and Engineering Research
Council of Canada.
K.B. is also supported by a Walter C. Sumner Memorial Fellowship.


\begin{thebibliography}{50}

\bibitem{link}
B.~Link,
Phys.\ Rev.\ Lett.\ {\bf 91}, 101101 (2003).

\bibitem{precession}
I.~H.~Stairs, A.~G.~Lyne, and S.~L.~Shemar,
Nature {\bf 406}, 484 (2000);
J.~Cordes, in {\it Planets Around Pulsars},
edited by Phillips, Thorsett, and Kulkarni
(1993), pp. 43-60.
T.~V.~Shabanova, A.~G.~Lyne, and U.~O.~Urama,
Astrophys. J. {\bf 552}, 321 (2001).

\bibitem{review} H. Heiselberg and V. Pandharipande,
arXiv:astro-ph/0003276.

\bibitem{typeI}
K.~B.~W.~Buckley, M.~A.~Metlitski and A.~R.~Zhitnitsky,
arXiv:astro-ph/0308148 (accepted to PRL).


\bibitem{witten}
E.~Witten,
Nucl.\ Phys.\ B {\bf 249}, 557 (1985).

\bibitem{defectsbook}
A.~Vilenkin and E.~P.~S.~Shellard, {\it Cosmic Strings and Other
Topological Defects}. Cambridge University Press, Cambridge, 2000.

\bibitem{volovik}
G.~E.~Volovik,
Phys.\ Rept.\  {\bf 351}, 195 (2001).
G.~E.~Volovik, {\it The Universe in a Helium Droplet}. Oxford
University Press, Oxford, (2003).

\bibitem{highTc}
S.~C.~Zhang, Science {\bf 275}, 1089 (1997).
D.~Arovas, J.~Berlinsky, C.~Kallin, and S.~C.~Zhang,
Phys.\ Rev.\ Lett.\ {\bf 79}, 2871 (1997);
K.~B.~W.~Buckley and A.~R.~Zhitnitsky,
Phys.\ Rev.\ B {\bf 67}, 174522 (2003).

\bibitem{krstrings}
D.~B.~Kaplan and S.~Reddy,
Phys.\ Rev.\ Lett. {\bf 88}, 132302, (2002).

\bibitem{superk}
K.~B.~W.~Buckley and A.~R.~Zhitnitsky,
JHEP {\bf 0208}, 013 (2002); 
K.~B.~W.~Buckley, M.~A.~Metlitski and A.~R.~Zhitnitsky,
Phys.\ Rev.\ D {\bf 68}, 105006, (2003).

\bibitem{BEC}
M.~A.~Metlitski and A.~R.~Zhitnitsky,
arXiv:cond-mat/0307559.
\bibitem{Bashkin}
A.~F.~Andreev and E.~P.~Bashkin,
Sov.Phys. JETP {\bf 42}, 164, (1976).
\bibitem{speight}
J.~M.~Speight,
Phys.\ Rev.\ D {\bf 55}, 3830 (1997).
\bibitem{mack}
R.~MacKenzie, M.~A.~Vachon and U.~F.~Wichoski,
Phys.\ Rev.\ D {\bf 67}, 105024 (2003).
\bibitem{Kramer}
L.~Kramer, Phys.\ Rev.\ B {\bf 3}, 3821 (1971).
\bibitem{COSLAB}
Third COSLAB workshop, July 10-16, 2003, Bilbao, Spain,
http://www.ehu.es/COSLAB

\bibitem{glitch1}
P.~W.~Anderson and N.~Itoh, Nature {\bf 256}, 25 (1975);
G.~Baym and C.~J.~Pethick, Ann. Rev. Astro. Astrophys. {\bf 17}, 415 (1979);
D.~Pines, in; Neutron stars: theory and observation of the Neutron Stars,
eds. J.~Alpar and D.~Pines, (Kluwer, Dordrecht, 1991) p. 57, and references
therein;
C.~J.~Pethick and D.~G.~Ravendall, Annu. Rev. Nucl. Part. Sci {\bf 45},
429 (1995).


\bibitem{glitchflux}
M.~Ruderman, T.~Zhu, and K.~Chen,
Astrophys. J. {\bf 492}, 267 (1998).

\bibitem{pinning}
B.~Link and C.~Cutler,
arXiv:astro-ph/0108281 (2001).

\bibitem{Baym}
G.~Baym, C.~Pethick, D.~Pines,
Nature {\bf 224}, 673 (1969).

\bibitem{LandauLam}
L.~D.~Landau,
Sov. Phys. JETP {\bf 7}, 371 (1937);
Zh.Eksp.Teor.Fiz. {\bf 13}, 377 (1943).


\bibitem{Sedrakian}
D.~M.~Sedrakian, A.~Sedrakian, G.~F.~Zharkov
Mon.\ Not.\ Roy.\ Astron.\ Soc. {\bf 290}, 203 (1997).



\end{thebibliography}
\end{document}